\documentclass[11pt,twoside]{article}
\usepackage{asp2006}

\usepackage{graphicx}
\usepackage{epsf}
\usepackage{psfig}
\usepackage{lscape}
\usepackage{natbib}

\begin{document}
\title{The AKARI Deep Fields: Early Results from\\ Multi-wavelength Follow-up Campaigns}

\author{Chris Sedgwick$^1$, Stephen Serjeant$^1$, Sandeep Sirothia$^2$, Sabyasachi Pal$^2$, Chris Pearson$^3$$^,$$^1$, Glenn White$^1$$^,$$^3$, Hideo Matsuhara$^4$, Shuji Matsuura$^4$, Mai Shirahata$^4$ and Sophia Khan$^5$ on behalf of the AKARI Deep Field North and South teams} 
  
\affil{$^1$The Open University $^2$NCRA,GMRT   $^3$Rutherford Appleton Laboratory $^4$ISAS, JAXA $^5$Pontificia Universidad Cat—lica, Chile}

\begin{abstract}
We present early results from our multi-wavelength follow-up campaigns of the AKARI Deep Fields at the North and South Ecliptic Poles. We summarize our campaigns in this poster paper, and present three early outcomes.  (a) Our AAOmega optical spectroscopy of the Deep Field South at the AAT has observed over 550 different targets, and our preliminary local luminosity function at 90$\mu$m from the first four hours of data is in good agreement with the predictions from Serjeant \& Harrison 2005.  (b) Our GMRT 610 MHz imaging in the Deep Field North has reached $\sim$30 $\mu$Jy RMS, making this among the deepest images at this frequency. Our 610 MHz source counts at $>$200 $\mu$Jy are the deepest ever derived at this frequency.  (c) Comparing our GMRT data with our 1.4 GHz WSRT data, we have found two examples of radio-loud AGN that may have more than one epoch of activity. 
\end{abstract}

\section{Introduction}

The AKARI Deep Field North (DFN) 2-26$\mu$m legacy survey is comprised of ultra-deep pointings with AKARI's Infrared Camera (IRC) and is AKARI's mid-infrared deep field. The AKARI Deep Field South (DFS), in contrast, is the premier far-infrared deep field. The covers 7 deg$^2$ in a contiguous slow-scan survey with AKARI's Far Infrared Surveyor (FIS) instrument, in four bands from 70-160$\mu$m, supplemented by AKARI IRC imaging over part of the field.

\section{Methods}

We are undertaking comprehensive multi-wavelength follow-ups to the AKARI Deep Fields: North - GMRT 610MHz radio, WSRT 1.4GHz radio, CFHT U-band imaging and SCUBA-2 Ultra Deep 450 and 850$\mu$m as part of Cosmology Legacy Survey; and South -  ATCA 1.4GHz radio, LABOCA 870 $\mu$m, AzTEC 1.1mm, CTIO R-band imaging and AAOmega optical-UV spectroscopy.

\begin{figure}[htbp]
\centering
\includegraphics[width=2.6in]{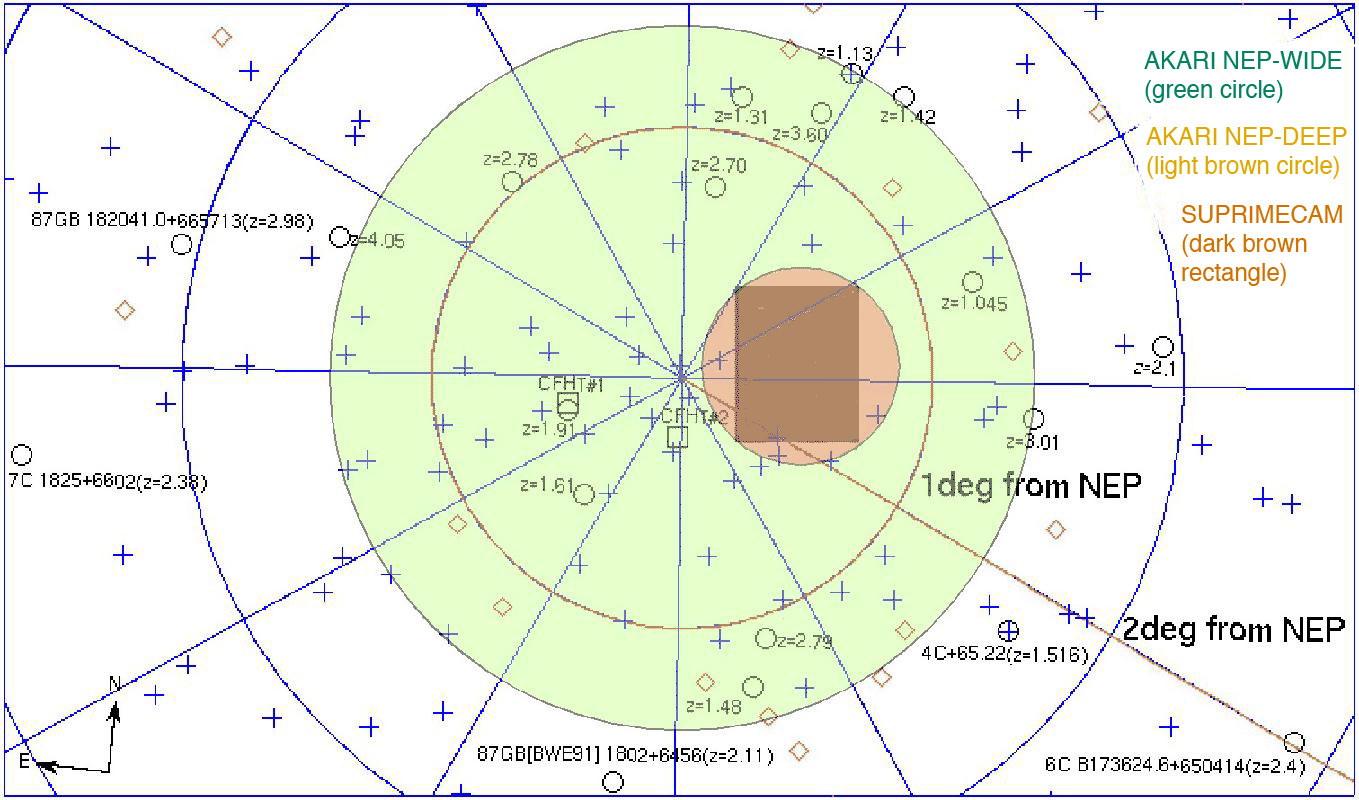}
\includegraphics[width=2.6in]{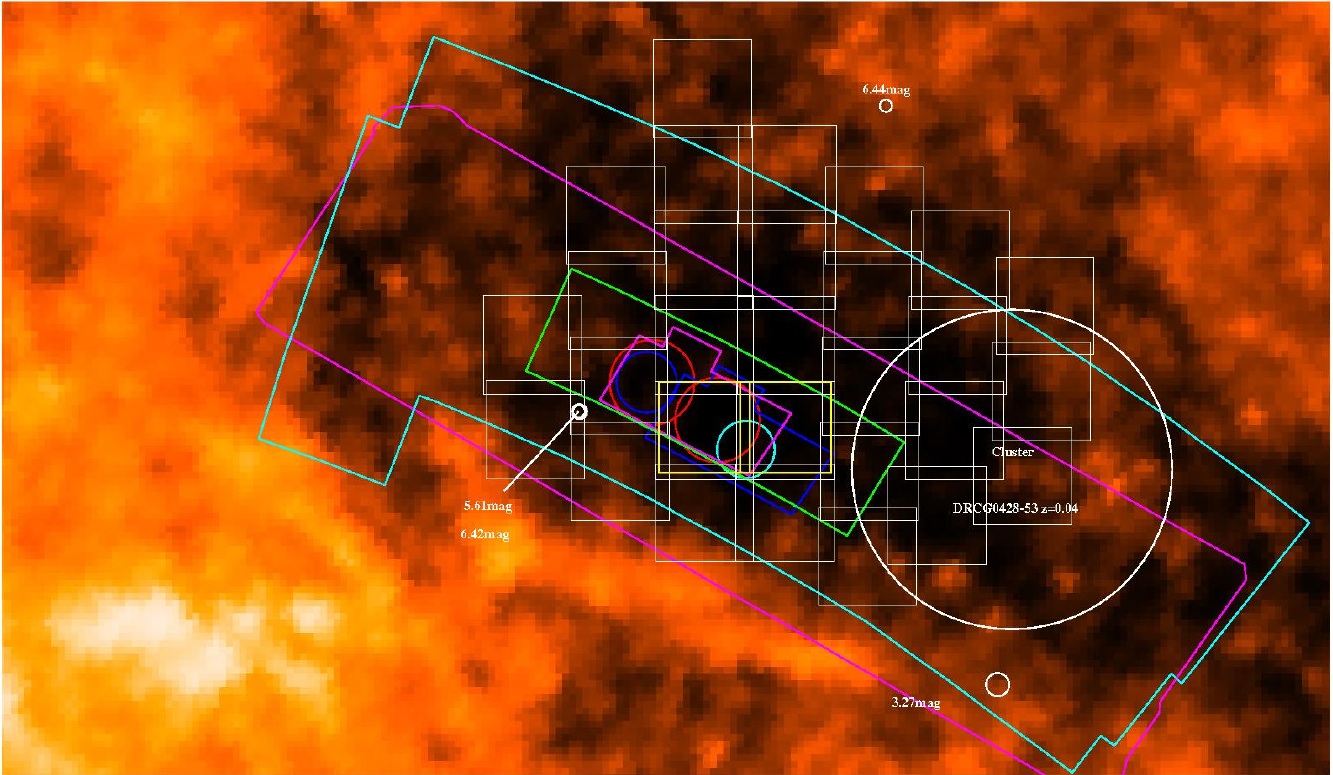}
\caption{Left: AKARI DFN (adapted from Matsuhara et al. 2006). AKARI NEP-Deep field is shown as a brown circle, AKARI NEP-Wide as a green shaded circle, and Suprimecam coverage as a brown rectangle. Right: AKARI DFS (50-180$\mu$m). 15-20 deg$^2$ fan-shaped survey using the FIS slow-scan mode, leading to 270 pointing observations. White squares: CTIO R band optical data; yellow squares: CTIO UVBRI optical data; big cyan box: AKARI FIS FIR data; big pink box: BLAST submm \& Spitzer IR data; red circles: ATCA radio; cyan circle: AzTEC mm; blue circle: LABOCA submm; small pink shape: IRC MIR data; small blue shape: IRC MIR data (courtesy of Mai Shirahata).}
\end{figure}

We have used AAOmega, the fibre-fed optical spectrograph at the Anglo-Australian Observatory, to obtain spectra of selected sources in the AKARI DFS. Data obtained October 2007 - November 2008 are currently being analyzed. 

We have undertaken a 610 MHz imaging survey using the Giant Metrewave Radio Telescope (GMRT) near Pune, India. The pilot study of 12 hours in 2007 (from which the diagram of source counts in Figure 5d was taken) was centred on the centre of the AKARI DFN and reached 30 $\mu$Jy RMS. A follow-up study of 24 hours has just been completed, slightly offset from this centre to increase the areal coverage. This survey represents one of the deepest images of the sky at this frequency. 

The preliminary 90$\mu$m local luminosity function in Figure 2 was estimated as follows.  (a) We used 70$\mu$m Spitzer counts in GOODS-N (Frayer et al. 2006) to estimate the AKARI completeness as a function of flux, scaling the fluxes by a factor of 0.673 on the grounds that the Pearson et al. 2001 source counts model has N(S70\textgreater0.0673Jy)= N(S90\textgreater0.1Jy). The completeness ranges from 97\% at 0.1Jy to 60\% at 28mJy. Future work will use a completeness estimated directly from the AKARI data. (b) We used APM B-magnitudes, but imposed a B\textless18 cut-off and assumed the completeness is not a function of B magnitude. This is justified on the basis that the AAOmega target selection at B\textless18 is essentially random. Further work will use the optically-fainter objects which extend to higher redshifts, including 90$\mu$m-selected objects identified on the basis of their mid-IR IRC identifications. (c) We estimated an effective sky coverage of the 2dF spectra using the fraction of B\textless14 objects which have redshifts. In order to remove large scale structure variations, we renormalised the effective area by the predicted numbers of \textgreater0.1Jy objects from the Spitzer counts (following the procedure adopted in the European Large Area ISO Survey, e.g. Serjeant et al. 2001). (d) We calculated the 1/V{\tiny max} luminosity function (Schmidt 1968) following the methodology in Serjeant et al. 2004, using the concordance cosmology of H{\tiny 0}=72 km/s/Mpc, $\Omega${\tiny M}=0.3, $\Omega${\tiny L}=0.7. The red line is the prediction from Serjeant \& Harrison 2005. At any luminosity log{\tiny 10}L we calculate the space density of objects with luminosities within log{\tiny 10}L$ \pm$0.25, shown as the hatched area in Figure 2.

\section{Results}

For the AKARI DFS, the AAOmega optical spectra have so far given us 246 redshifts for AKARI sources.  Of these, 60 have z\textless0.1 (mostly FIS sources) and 28 have  z\textgreater0.5 (mostly IRC sources). A preliminary Luminosity Function based on some of these early results is shown in Figure 2. Figures 3 and 4 show examples of further results for AKARI DFS. For the AKARI DFN, Figure 5 shows a preliminary differential source count based on GMRT data, Figure 6 shows radio and optical comparisons for parts of the field, and Figure 7 compares GMRT and WSRT images of  AGN that may have two epochs of radio activity.\\

\begin{figure}[htbp]
\centering
\includegraphics[width=4.3in]{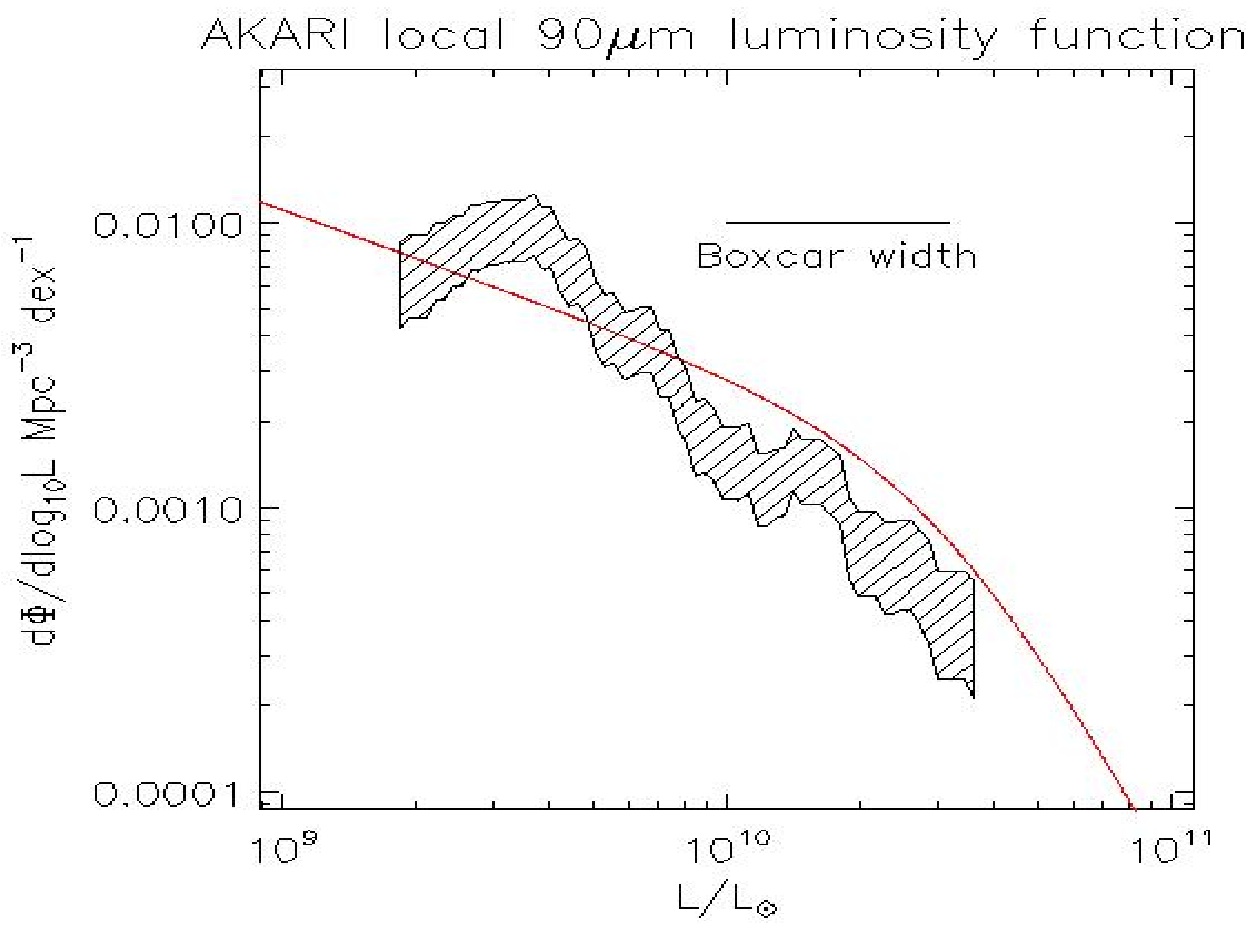}
\caption{Preliminary  90$\mu$m Local Luminosity Function. With 4 AAOmega runs in the DFS, we targeted 550 objects in the first 0.5 nights, from the IRC, FIS, WSRT, LABOCA \& AzTEC catalogues.  At any luminosity log{\tiny 10}L we calculate the space density of objects with luminosities within log{\tiny 10}L$ \pm$0.25 (hatched area). Red line is prediction from Serjeant \& Harrison 2005.}
\end{figure}

\begin{figure}[htbp]
\centering
\includegraphics[width=1.2in]{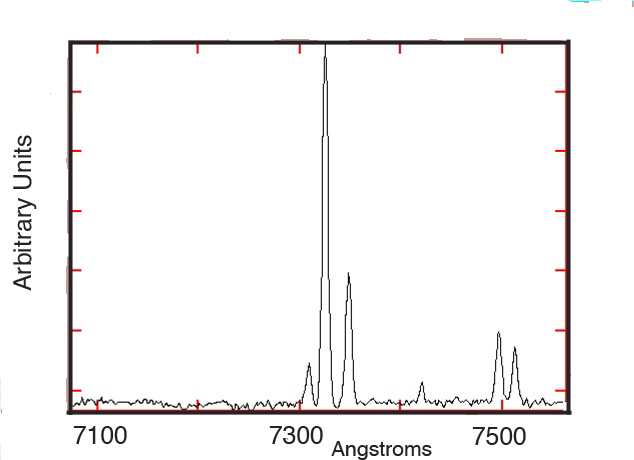}
\includegraphics[width=1.2in]{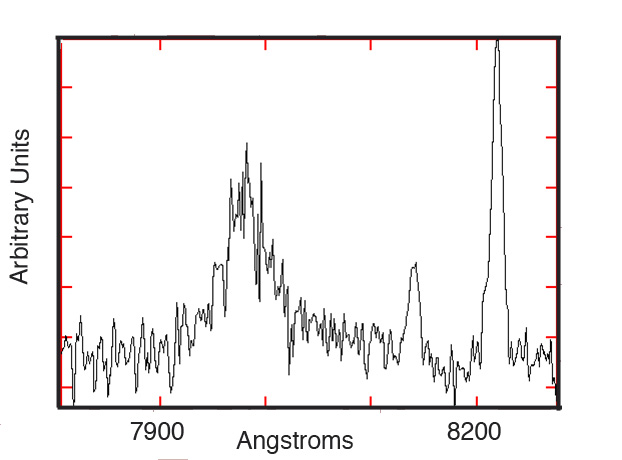}
\includegraphics[width=0.85in]{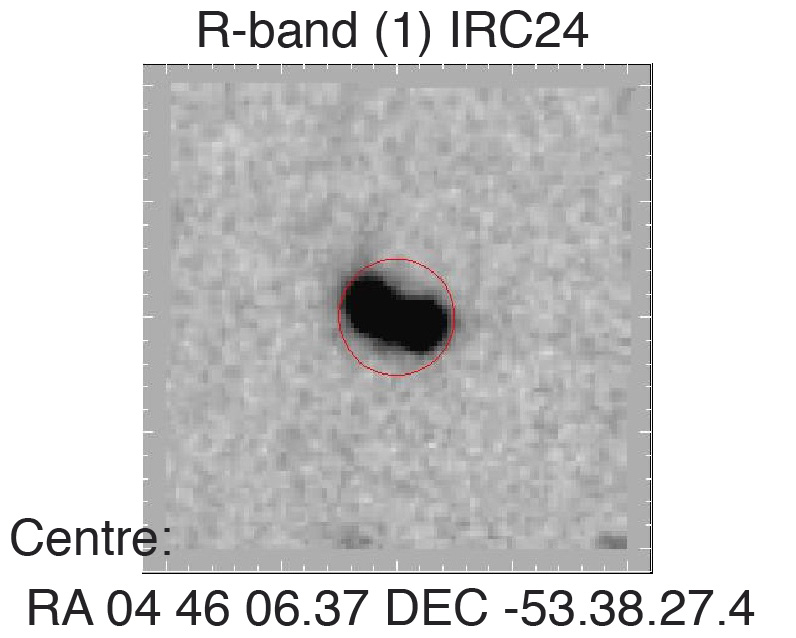}
\includegraphics[width=0.85in]{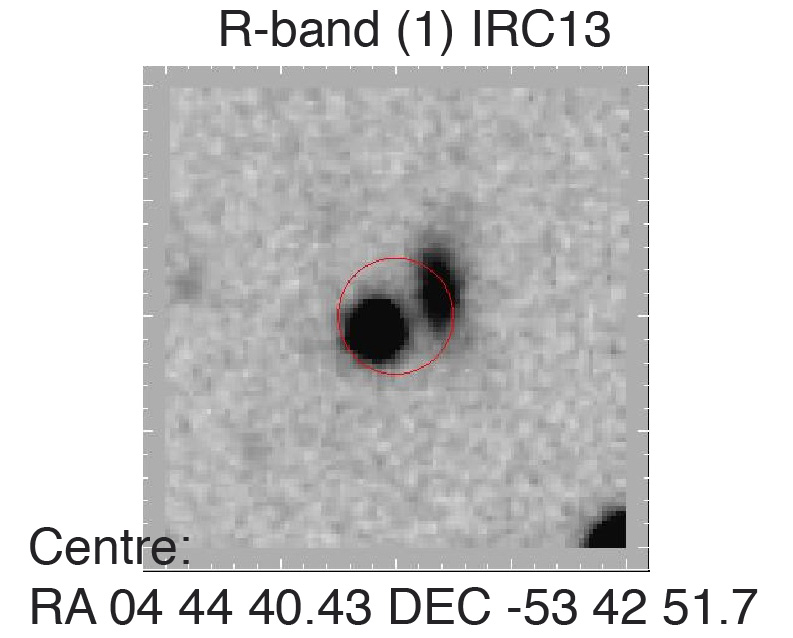}
\includegraphics[width=0.85in]{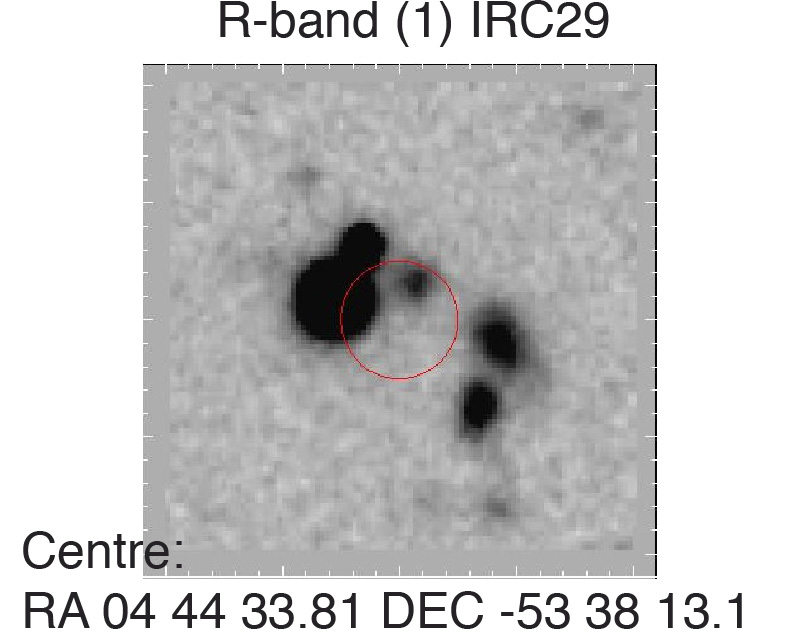}
\caption{(a) FIS552 spectrum showing H$\alpha$/[NII]$\lambda$$\lambda$6548,6563,6583\& [SII]$\lambda$$\lambda$6716, 6730 lines. (b) IRC3 spectrum showing H$\beta$ /[OIII] $\lambda$$\lambda$4861, 4958,5007 lines at z=0.64. (c)-(e): CTIO R-band images IRC24,13 \&29.}
\end{figure}

\begin{figure}[htbp]
\centering
\includegraphics[width=1.2in]{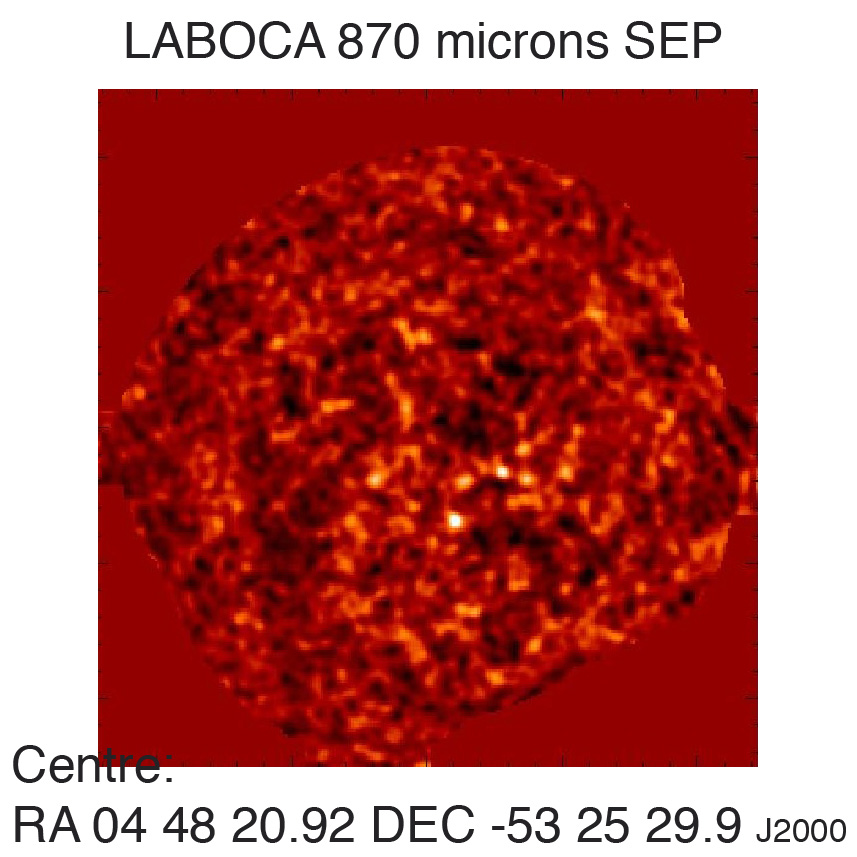}
\includegraphics[width=3.9in]{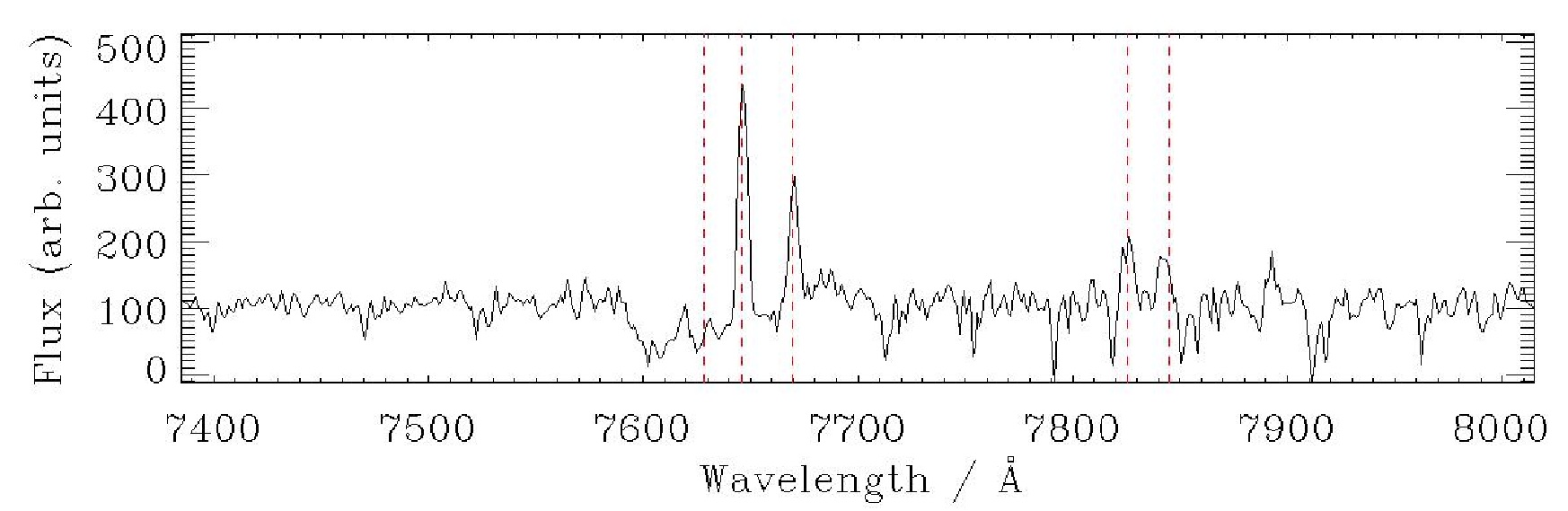}
\caption{(Left) 870$mu$m LABOCA signal-to-noise image taken in the AKARI DFS. Black is -3$\sigma$ and white is +5$\sigma$. The depth is approximately 4$\sigma$ 8mJu. Further details in Khan et al 2009 (in preparation). (Right) We observed several AzTEC sources using AAOmega, obtaining six redshifts. This example shows one  at z=0.165 with redshifted H$\alpha$, [NII] and [SII] lines.}
\end{figure}

\begin{figure}[ht]
\centering
\includegraphics[width=1.2in]{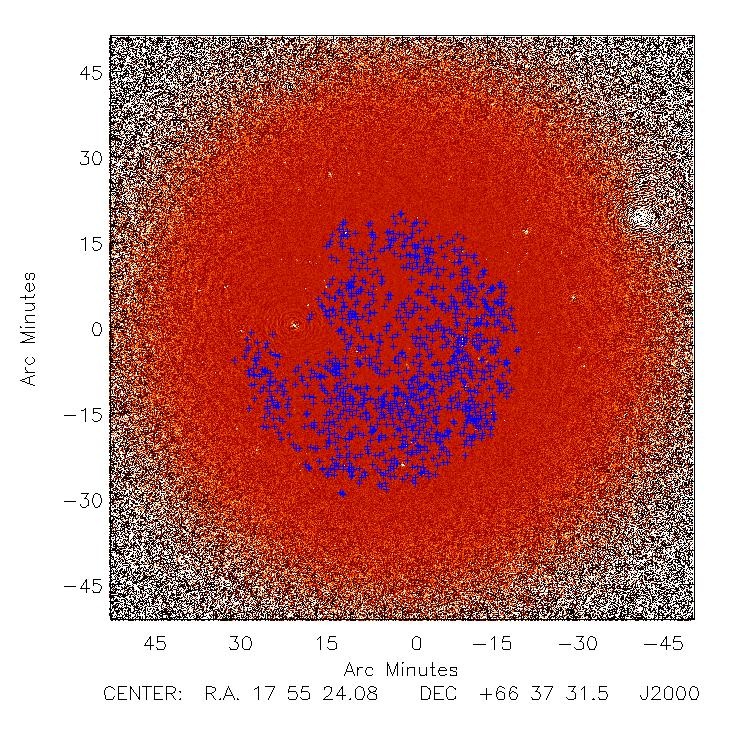}
\includegraphics[width=1.1in]{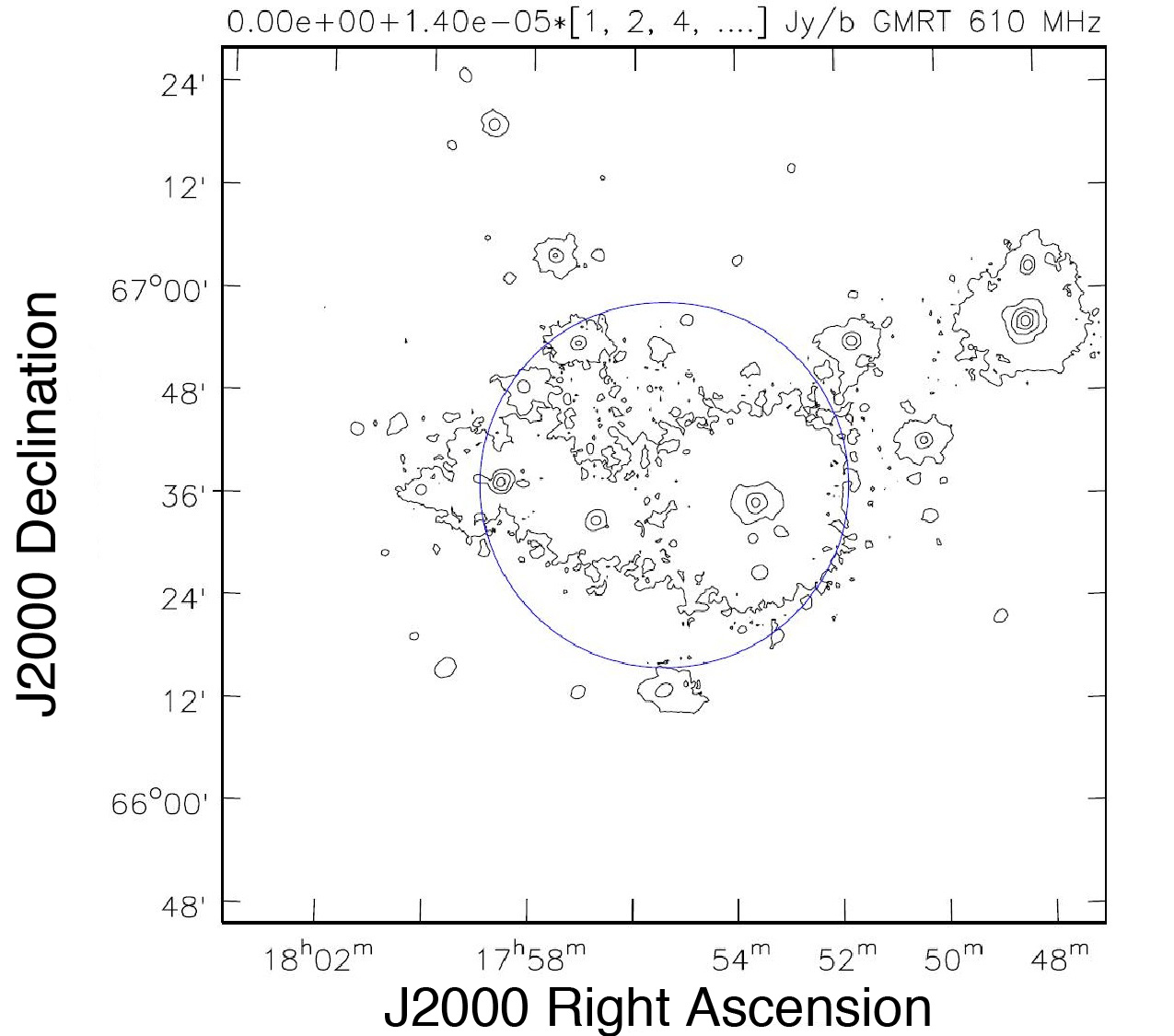}
\includegraphics[width=1.2in]{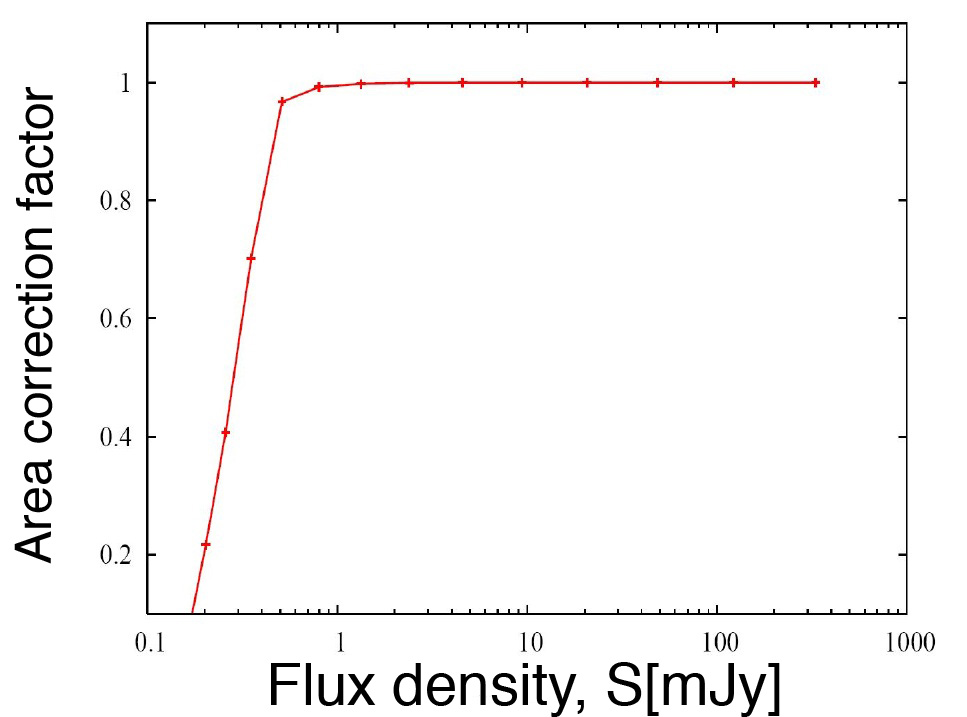}
\includegraphics[width=1.6in]{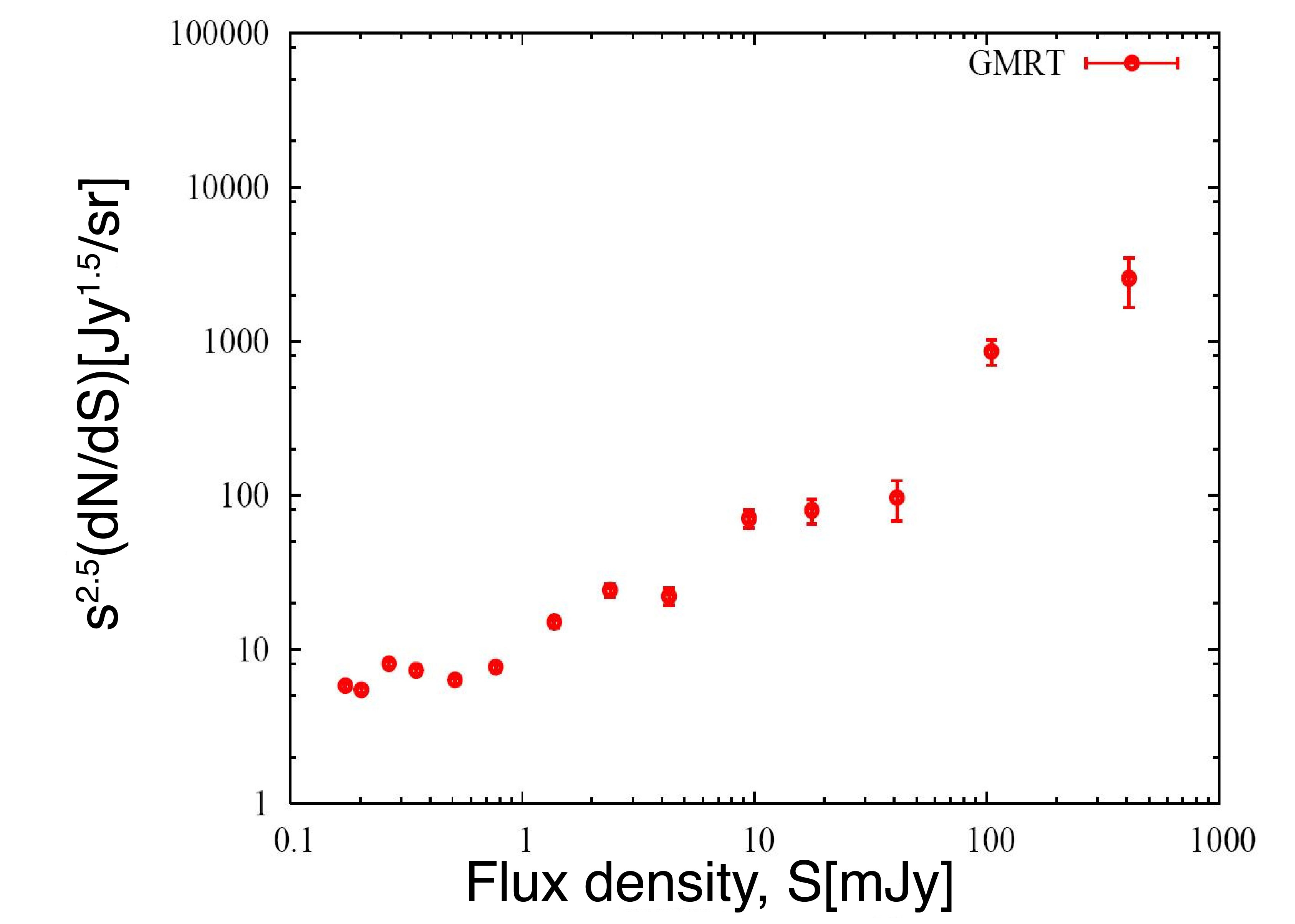}
\caption{(a) GMRT 610 MHz image, scaled from -1mJy (black) to +1mJy (white). The AKARI DFN all-band detected sources have been over-plotted as blue crosses (b) The variation of RMS noise across the image before primary beam correction. Contour levels in units of Jy per beam are represented by mean + RMS x n where n is an integer. Negative contours shown as dashed lines (c) Visibility area as a function of radio flux density (d) Euclidean normalized differential source counts (Sirothia et al 2009 in preparation).}
\end{figure}

\begin{figure}[!htbp]
\centering
\includegraphics[width=3.3in]{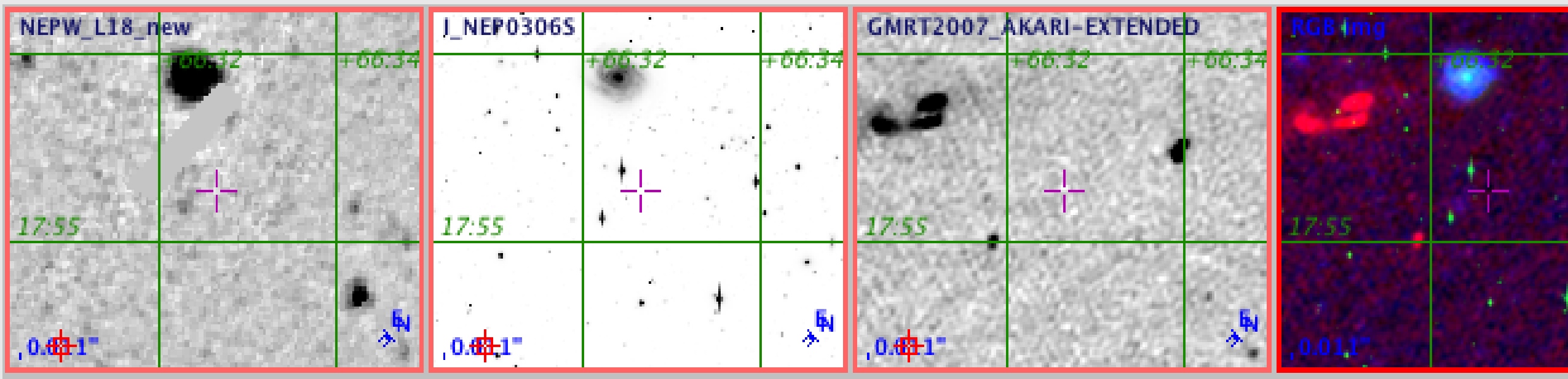}
\includegraphics[width=3.3in]{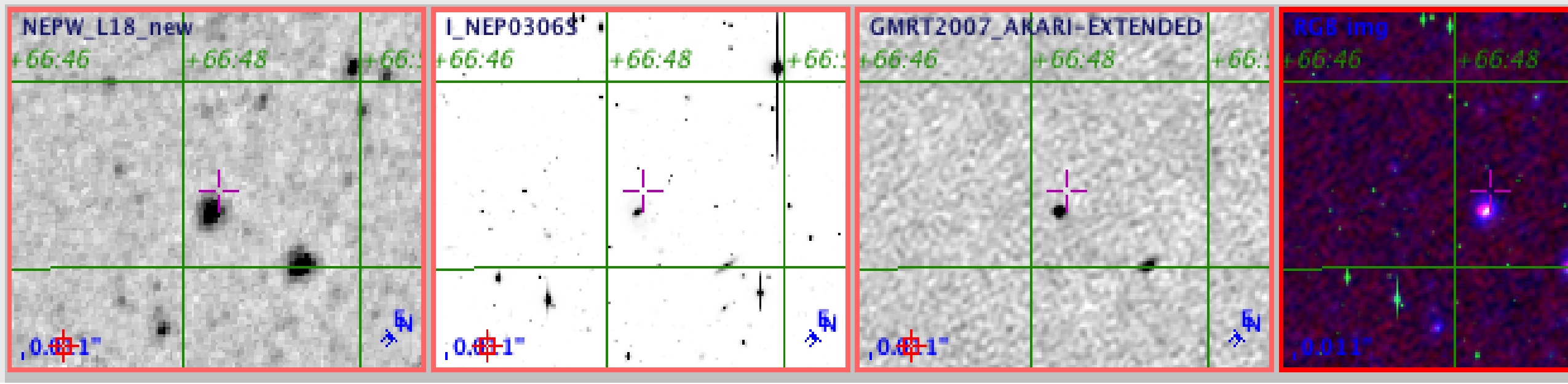}
\caption{Two radio and optical comparisons to parts of the AKARI  DFN region. In each case from left: AKARI IRC 18$\mu$m, Subaru Suprimecam I-band, GMRT 610 MHz,  RGB Composite (Red=GMRT, Green=I-band Subaru, Blue=AKARI 18$\mu$m) (Sedgwick et al 2009 in preparation).}
\end{figure}

\begin{figure}[!htbp]
\centering
\includegraphics[height=1.1in]{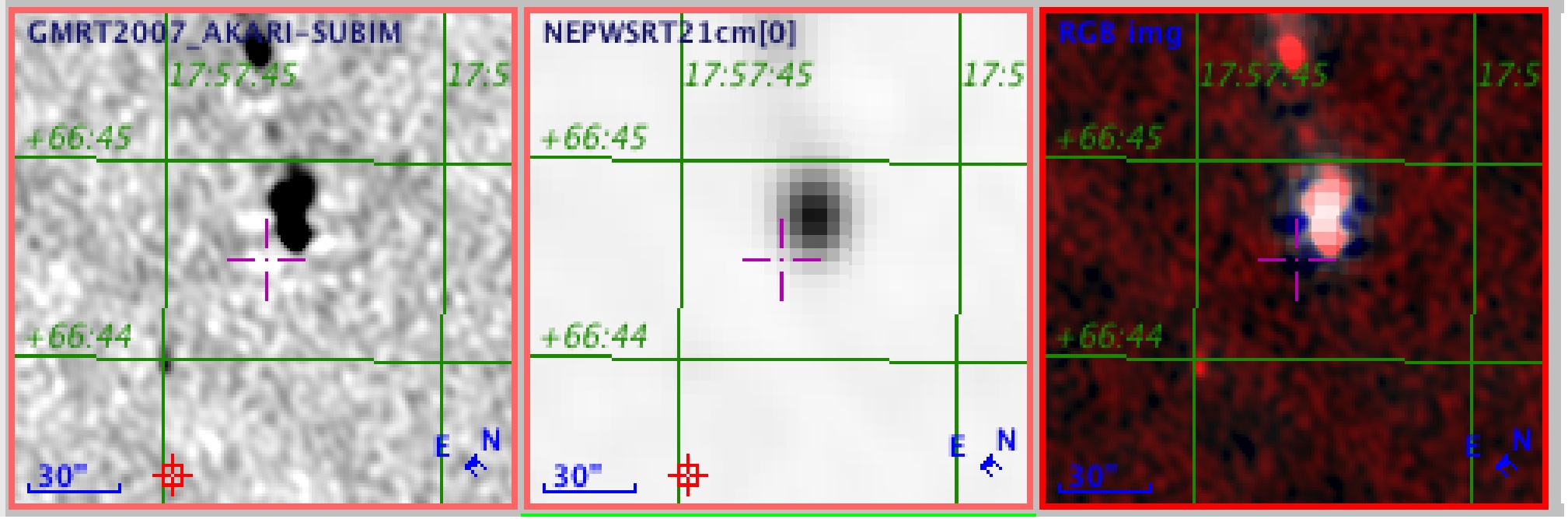}
\includegraphics[height=1.13in]{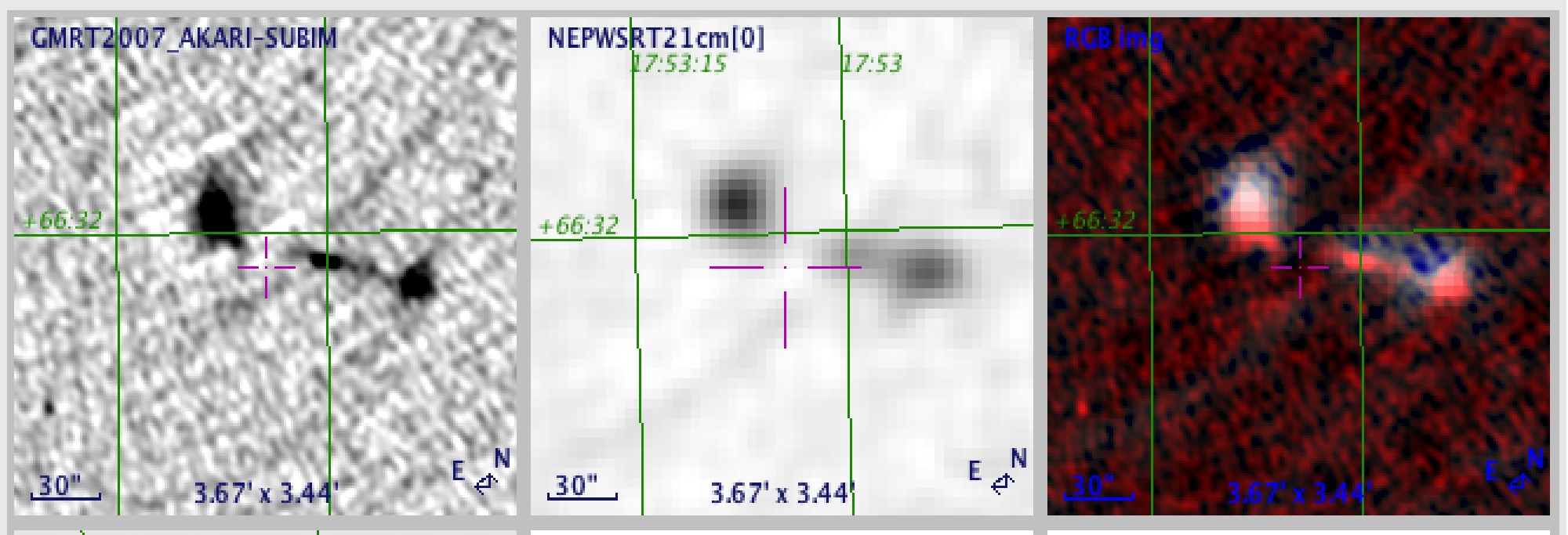}
\caption{Candidate examples of AGN that may have two epochs of radio activity: (left) 17 57 40 +66 45 00 (right) 17 53 00 +66 31 40. In each case, left: GMRT 610MHz, middle: Westerbork Radio Synthesis Telescope 1.4GHz, right: Combined (Red=GMRT, Blue=WSRT). These are the subject of planned follow-up radio observations (Pal et al 2009 in preparation).}
\end{figure}

\acknowledgements 
This work has been funded in part by STFC (grant  PP/D002400/1), Royal Society (2006/R4-IJP) \& Sasakawa Foundation (3108).\\\\\\\\\

\end{document}